\documentstyle[12pt]{article}
\renewcommand
\baselinestretch{1.4}

\begin{document}

\hfill{KEK-TH-406}

\hfill{UM-P-94/80}

\hfill{RCHEP 94/24}
\vskip .5cm
\begin{center}

{\LARGE \bf
BRS Symmetry in Connes' Non-commutative Geometry}

\renewcommand
\baselinestretch{0.8}\vspace{4mm}
{\sc B. E. Hanlon}
\\
{\it
KEK Theory Group \\
Tsukuba \\
Ibaraki 305 \\
Japan}
\\ and \\
{\sc G. C. Joshi}
 \\
{\it
Research Centre for High Energy Physics \\
School of Physics \\
University of Melbourne \\
Parkville, Victoria 3052 \\
Australia}

\renewcommand
\baselinestretch{1.4}

\begin{abstract}
We extend the BRS and anti-BRS symmetry to the two point space of Connes'
non-commutative model building scheme. The constraint relations are derived
and the quantum Lagrangian constructed. We find that the quantum Lagrangian
can be written as a functional of the curvature
for symmetric gauges with the BRS, anti-BRS
auxiliary field finding a geometrical interepretation as the extension of the
Higgs scalar.
\end{abstract}

\end{center}

\vspace{0.5cm}


\newpage
\section{Introduction}

Non-commutative geometry has emerged as a promising model building
prescription providing a possible underlying structure to the appearance of
Higgs scalars with point-like interactions and quartic curvatures. As the
name suggests this is achieved by generealizing the underlying notion of
geometry as applied to particle physics. This generalization is supported by
the Gelfand-Naimark theory: let $\cal A$ be a $C^{\ast}$-algebra with unity,
then if $\cal A$ is commutative it will be isomorphic to the algebra of all
continuous complex-valued functions $C^{\infty}(X)$, defined on a compact
topological space, $X$. Thus rather than the manifold itself one addresses the
algebra of smooth functions defined over it, which is equivalent. A
differential calculus is then constructed on $C^{\infty}(X)$. This provides a
very convenient way to generalize the topological space, $X$, in a sense
linearising the description of complicated or ``badly behaved" spaces.

There is more than one approach to implementing this geometric view. One
perspective, developed by Dubois-Violette, Kerner and Modore, extends the set
of complex smooth functions over space-time to include complex matrices i.e.
$C^{\infty}(M)\rightarrow C^{\infty}(M)\otimes M_{n}({\bf C})$; $M$ being
space-time and $M_{n}({\bf C})$ the set of $n\times n$ complex matrices{\cite
A}. It
is in this sense that the geometric prescription is non-commutative; see also
Balakrishna, G\"ursey and Wali{\cite B}. The important aspect here is that the
differential calculus is developed on the entire algebra so one does not
simply yield matrix valued forms. Indeed Higgs scalars emerge as generalized
one forms valued in the derivation algebra of $M_{n}({\bf C})$. This is
reminiscent of BRS analysis in normal gauge theory, indicating perhaps a
deeper underlying assosciation.

The non-commutative model building scheme which will be of particular
interest in this paper is that developed by Connes{\cite C}
; see also Connes and Lott{\cite {D,E}}
as well as Chamseddine, Felder and Fr\"ohlich{\cite F}
who reformulated this approach
to include GUT models. Connes generalized the geometric prescription by
extending the algebra of smooth functions to include a two point space, thus
e.g. $C^{\infty}(M)\rightarrow C^{\infty}(M)\oplus C^{\infty}(M)$. Gauge fields
arise as appropriately defined fibre bundles on each copy of space-time while
Higgs scalars appear as connections between these copies. In this way the
symmetry breaking scale receives a geometrical interpretation being just the
inverse distance between space-times. This construction places severe
restrictions on the models which can be constructed. For instance,
flavour chirality is essential. Furthermore, survival of the Higgs
potential demands the existence of multiple, non-degenerate, fermionic
families. This is an intriguing correspondence with phenomenology{\cite G}. It
appears
also that constraints upon the Higgs and top quark masses appear at the
classical level{\cite H}. This is a natural consequence of entering fermionic
data
into the model as a starting point towards reproducing observed
behaviour.

The standard model constructed by Connes and Lott is a remarkable success of
this model building prescription{\cite E}. Nevertheless a quantum theory is
still
lacking. This state of affairs is not unreasonable given the intrinsically
new setting of this approach. Indeed connections with quantum
theory are now emerging, perhaps suggesting that the non-commutative settings
are fundamentally quantum mechanical{\cite I}. On a less ambitious level,
however, it
appears that the constraints imposed by non-commutative geometry do not
survive quantum corrections{\cite J}. This is a common symptom of such
``Kaluza-Klein
like" model building schemes (see also the connection with coset space
dimensional reduction{\cite K}). However, given the novelty of this geometric
description, it is not unreasonable that an additional symmetry or some
more exotic mechanism exists in which quantum corrections are consistent{\cite
L}. In
this context an intrinsic quantum mechanical connection is indeed compelling.

The failure of the non-commutative model building constraints to survive
quantum corrections was demonstrated in only the simplest possible abelian
model (the standard model evolution was shown, tentatively, to be slow{\cite
J}).
Given that the quantum connection has not, nevertheless, been
satisfactorily resolved we wish in this paper to extend to non-abelian models
and consider the non-commutative implementation of BRS symmetry in Connes'
model building scheme. Independently of providing a possible framework for
quantizing such models we seek to generalize the gauge fixing mechanism at
the classical level. This is made possible by the geometric origin of the BRS
and anti-BRS constraints{\cite M}. Since in non-commutative geometry the Higgs
scalar
appears on the same level as gauge fields we have a conceptually simple means
by which matter fields may be included in the notion of BRS symmetry. This is
compelling if one recalls the important role that Higgs scalars play in
interesting soultions of Yang-Mills fields, such as monoploes. We find that the
extended geometric setting of non-commutative geometry provides a framework
in which a unified description may be derived with a more adequate
geometrical interpretation of the BRS/anti-BRS scalar emerging.

\section{Non-commutative Gauge Theory}

We will briefly overview the model building prescription of non-commutative
gauge
theory to set the mathematical formalism. While new perspectives to quantum
theory are
expanding the motivation for this approach{\cite {I,K}} the original motivation
was
geometric and
we shall introduce the scheme in this setting.

To generalize the Riemannian metric the notion of geodesic distance
must be consistently incorporated and this is encoded in the concept
of a K-cycle. A K-cycle over the involutive algebra, $\cal B$ say, is
a $\ast$-action
of $\cal B$ by bounded operators on a Hilbert space $\cal
H$, denoted by $\rho$, and a possibly unbounded, self adjoint, operator
$D$, denoted Dirac operator, such that $[D,\rho (f)]$ is a bounded operator
$\forall f\in {\cal B}$ and $(1+D^{2})^{-1}$ is compact. Let $X$ be a
compact Riemannian spin manifold, $\cal F$ the algebra of functions on
$X$ and $({\cal H}_{\cal F},{\not \! \partial})$ the Dirac K-cycle
with ${\cal H}_{\cal F} = L^{2}(x,{\sqrt g}d^{d}x)$ of $\cal F$.
Denote by $\gamma_{5}$ the fifth anticommuting Dirac gamma matrix, the
chirality operator, defining a $Z_{2}$ grading on ${\cal H}_{\cal F}$.
Similarly a K-cycle can be defined on the discreet set representing
the internal space. Let $\cal A$ be given by ${\cal A} =
M_{n}({\bf C})\oplus M_{p}({\bf C})\oplus ....$
corresponding to the Hilbert
spaces ${\bf C}^{n},{\bf C}^{p},....$
respectively. (Alternatively, one may define an n-point space ${\cal A}=
C^{\infty}(X)\oplus C^{\infty}(X)\oplus ...$ and introduce an appropriate
vector
bundle ${\cal E} = e{\cal A}^{p}$, $p\in {\bf Z}$, where $e$ is a projection.
The
algebra $\cal A$ and the vector bundle $\cal E$ then act together to define the
required gauge groups on each copy of space-time{\cite E}).
For a two point space the
Dirac operator takes the form
\begin{equation}
D =
\left (
\begin{array}{cc}
 0 \;\;\; M  \\
 M ^{\ast}\;\; 0
\end{array}
\right )
\;\; ,
\end{equation}
$M$ being a mass matrix of size dim${\cal H}_{L}\times$dim${\cal
H}_{R}$, where ${\cal H}={\cal H}_{L}\oplus {\cal H}_{R}$ decomposes
under the action of a suitable chirality operator
\begin{equation}
\chi =
\left (
\begin{array}{cc}
 1_{L} \;\;\;\; 0\;  \\
 0 \; {-1_{R}}
\end{array}
\right )
\;\; .
\end{equation}
For a product geometry ${\cal A}_{t}={\cal F}\otimes {\cal A}$ a
product K-cycle is naturally defined with the generalized Dirac
operator
\begin{equation}
D_{t}=
{\not \! \partial}\otimes 1 +\gamma_{5}\otimes D \;\; .
\end{equation}
For any $C^{\ast}$-algebra a K-cycle will define a metric $d$ on the
state space of $\cal B$ by
\begin{equation}
d(p,q)={\rm sup}\{ |p(f)-q(f)|:f\in {\cal B}, ||[D,f]|| \leq 1\} \;\; .
\end{equation}
Recall that now the points $p$ and $q$ are states on the algebra so that
$p(f)\equiv f(p)$.

To make contact with gauge theories a differential algebra must be
constructed on $\cal B$. The space of all differential forms
${\hat{\Omega}}^{\ast}({\cal B})=\bigoplus_{p\in N}{\hat{\Omega}}^{p}
({\cal B})$ is a graded differential algebra equipped with a
differential operator $\delta$ such that
\begin{equation}
\delta :{\hat{\Omega}}^{p}({\cal B})\rightarrow
{\hat{\Omega}}^{p+1}({\cal B}) \;\; ,
\end{equation}
along with nilpotency
\begin{equation}
\delta^{2}=0 \;\; .
\end{equation}
The space of $p$-forms ${\hat{\Omega}}^{p}({\cal B})$ is generated by
finite sums of terms of the form
\begin{equation}
{\hat{\Omega}}^{p}({\cal B}) =\{ \sum_{j} a_{0}^{j}\delta
a_{1}^{j}.....\delta a_{p}^{j}, a_{p}^{j}\in {\cal B}\} \;\; ,
\end{equation}
which follows from the relations
\begin{eqnarray}
\delta 1 = 0 \hspace{1.25cm}\nonumber \\
\delta (ab)=(\delta a)b + a\delta b \;\; ,
\end{eqnarray}
with the differentail $\delta$ defined by
\begin{equation}
\delta (a_{0}\delta a_{1}.....\delta a_{p}) = \delta
a_{0}\delta a_{1}....\delta a_{p} \;\; .
\end{equation}

Extending the representation of $\cal B$ on $\cal H$ to its universal
differential envelope ${\hat{\Omega}}^{\ast}({\cal B})$ is achieved
via the map
\begin{equation}
\pi : {\hat{\Omega}}^{\ast}({\cal B})\rightarrow B({\cal H}) \;\; ,
\end{equation}
$B({\cal H})$ being the algebra of bounded operators on $\cal H$
defined by
\begin{equation}
\pi (a_{0}\delta a_{1}.....\delta a_{p})=\rho (a_{0})[D,\rho
(a_{1})]....[D,\rho (a_{p})] \;\; .
\end{equation}
It is this representation on Hilbert space, as a way of connecting with our
usual
notions of space-time vectors and scalars, which distinguishes the Connes-Lott
model
building scheme{\cite E}.
The crucial aspect which must be considered, however, is that the
representation
$\pi$ is ambiguous, with the correct space of forms actually given by
\begin{equation}
{\Omega}^{\ast}({\cal B}) = {\hat{\Omega}}^{\ast}({\cal B})/J \;\; ,
\end{equation}
$J$ being given by the differential ideal{\cite{D,G}}
\begin{equation}
J={\rm ker}\pi +\delta {\rm ker}\pi = \bigoplus_{p} J^{p} \;\;,
\end{equation}
where
\begin{equation}
J^{p} = ({\rm ker}\pi )^{p} + \delta ({\rm ker}\pi )^{p-1} \;\; .
\end{equation}
Degree by degree the correct space of forms becomes
\begin{eqnarray}
\Omega^{0}({\cal B}) ={\hat{\Omega}}^{0}({\cal B})\cong \rho ({\cal
B}) \;\; , \hspace{.7cm}\nonumber \\
\Omega^{1}({\cal B}) ={\hat{\Omega}}^{1}({\cal B})/({\rm ker} \pi )^{1}
\cong \pi ({\hat{\Omega}}^{1}({\cal B})) \nonumber
\end{eqnarray}
and for degree $p\geq 2$
\begin{equation}
\Omega^{p}({\cal B}) \cong\pi ({\hat{\Omega}}^{p}({\cal B}))/ \pi (\delta
({\rm ker} \pi )^{p-1} ) \;\; .
\end{equation}
When ${\cal B} = {\cal F}$, ${\Omega}^{\ast}({\cal B})$ recovers
deRham's differential algebra of differential forms on $X$.

The simplest means by which the quotient may be considered as a
subspace is in the prescence of a scalar product. This is naturally
defined on the internal space by the trace on matrices. In the
infinite dimensional case the inner product is defined by the Dixmier
trace
\begin{equation}
{\rm tr}_{\rm w} (Q|{\not \! \partial}|^{-d})={\lim_{N\rightarrow
\infty}} {1\over\log N} \sum_{n=1}^{N}\lambda_{n} \;\; ,
\end{equation}
where $Q$ is a bound operator on ${\cal H}_{\cal F}$, $d=$dim$X$ and
$\lambda_{n}$ are eigenvalues of $Q|{\not \! \partial}|^{-d}$ arranged
in a decreasing sequence discarding the Dirac zero modes. By
correspondence with the deRham complex on $X$ this reduces to the
usual scalar product on $X$,
\begin{equation}
(\phi ,\psi ) = 1/8\pi^{2} \int_{X} \phi^{\ast}\ast\psi \;\;\;\;\;\;
\phi,\psi\in \Omega^{p} (X)
\end {equation}
incorporating the Hodge star. Since the fermionic fields are the
fundamental fields the spinor action can be simply written down.

\section{The Geometry of BRS Symmetry}

As is well known, difficulties arise when covariantly quantizing non-abelian
gauge theories from unwanted contributions to the gluon propogator.
Transversality and unitarity are spoilt in closed loop diagrams from the
longitudinal part of the propogator. To overcome this it is first necessary
to introduce a constant notion of transversality on the gauge field and this
is implemented by gauge fixing (up to problems arising from the
Gribov ambiguity).
Since differing gauges need not be smoothly
connected it becomes necessary to restrict to an appropriate region of
configuration space by choosing a particular representative in some
equivalence class of gauge related connections. In the functional integral
representation the Jacobian of this gauge fixing term can be written in terms
of a set of fermionic scalars, known as ghosts{\cite N}. Since loop diagrams
involving
such ghosts will introduce factors of (-1) due to their fermionic nature they
will ensure the cancellation of unitarity violating terms in the perturbation
expansion. (In the abelian case the Ward identities are satisfied without the
need for ghosts). The gauge fixing term results in the loss of gauge
invariance for the new ``quantum Lagrangian" now consisting of the original
terms plus the gauge fixing and ghost contributions. However, a new global
symmetry can be defined with all the consequences of gauge invariance, this
is the BRS symmetry.

The geometrical origin of the BRS (as well as anti-BRS) invariance was
demonstrated by Baulieu and Thierry-Mieg by reversing the construction of
Yang-Mills theories{\cite M}. Gauge fields and ghosts are now introduced as the
fundamental independent fields. This is motivated by the geometrical
description of gauge theories in terms of principal fibre bundles, which
consist of a base space (space-time with co-ordinates $x$) and fibres
corresponding to local copies of the gauge group (with co-ordinates $y$). A
matter field in this $(x,y)$ space takes the form
\begin{equation}
\tilde{\psi} (x,y) = {\rm exp}
(iy^{m}T^{\alpha}_{m\beta})\psi^{\beta} (x) \;\; ,
\end{equation}
$T_{m}$ being the generators of the Lie group. This describes our usual
notion of gauge transformation. The extension is made when one generalizes
the gauge field one-form $A_{\mu} (x)dx^{\mu}$ to the $(x,y)$ space
\begin{equation}
\tilde{A^{a}}(x,y)= A_{\mu}^{a}(x,y)dx^{\mu} + C_{m}^{a}(x,y)dy^{m} \;\; ,
\end{equation}
$C_{m}^{a}(x,y)$ being a scalar field with $dx^{\mu}, dy^{m}$ spanning the
cotangent space of the fibre bundle. The ghost field is identified with
$C_{m}^{a}(x,y)dy^{m}$ which anticommute by virtue of being differential
forms. The anti-BRS ghost field can be similarly introduced by constructing a
``double" principal bundle with co-ordinates $(x,y,{\overline y})$. This is
isomorphic to the product of space-time by two copies of the gauge group.
However, this is not to be confused with the two point space of Connes'
construction{\cite E} as there is no notion of geodesic distance between these
copies.
The generalized gauge field now becomes
\begin{equation}
\tilde{A^{a}}(x,y,{\overline y})= A_{\mu}^{a}(x,y,{\overline y})dx^{\mu} +
C^{a}_{m}(x,y,{\overline y})dy^{m} +
{\overline C}^{a}_{m}(x,y,{\overline y})d{\overline y}^{m} \;\; .
\end{equation}
BRS and anti-BRS equations are now introduced as geometrical constraints
arising from constructing the generalized curvature
\begin{equation}
\tilde{F^{a}} = \tilde{d}\tilde{A^{a}}+{1/2}[\tilde{A},\tilde{A} ]^{a} \;\; .
\end{equation}
Imposing the Cartan-Maurer condition insures compatability of the fibration
with parallel transport, restricting the curvature to be proportional to
$dx^{\mu}\wedge dx^{\nu}$. The remaing terms, such as $dy^{m}\wedge dx^{\mu}$,
must cancel, thus yielding the required equations.

The association of this approach to the non-commutative model building scheme
of Dubois-Violette et al.{\cite A} should now be clear. Indeed this was
directly
exploited by Balakrishna et al.{\cite B}. The difference lies in the
interpretation of
these new scalar fields as being bosonic or fermionic. Extending Connes'
approach then will effectively constitute employing both model building
directions, that of Connes{\cite {C,D,E}} and that of Dubois-Violette et al.
{\cite A}.This has in
fact been used before to yield sufficiently diverse sets of Higgs scalars to
accomodate required symmetry breaking patterns{\cite O}. Importantly, it
appears that
in exploring BRS symmetry we will be exploiting the notion of ``internal
space" to its fullest limits in deriving a suitable model.

\section{Application to Parallel Space-Times}

That interesting generalizations may emerge from applying these notions to
Connes' construction is amply demonstrated by the richness of structure when
cohomological considerations are directed to non-commutative geometry{\cite P}.
As
the previous construction suggests, we wish to extend our algebraic
considerations onto the group manifolds on each copy of space-time.
This has already been explored by Watamura for quantum groups in the case
of one space-time{\cite Q}. This sets the mathematical tone, where on each copy
of space-time we will restrict ourselves to the classical groups,
each extended copy now being
described by the algebra
\begin{equation}
C^{\infty}(X\otimes G)\cong C^{\infty}(X)\otimes C^{\infty}(G) \;\; ,
\end{equation}
the isomorphism arising
due to the local triviality of the principal fibre bundle.
Ghosts will now appear due to $C^{\infty}(G)$, $G$ being a compact
unitary group.
Our analysis will differ in that a set of parallel space-times will also
introduce scalar degrees of freedom as connections between these copies.
This brings us to the interesting role that matter fields will now play
in our analysis. The idea is very simple. As with other model building
schemes which exploit ``internal'' structure, the Higgs scalar and gauge
fields are recognized as originating from a single underlying principal,
i.e. they are unified. We propose then to extend the scalar contributions
in an analogous manner to that of the gauge fields in (20). Thus
\begin{equation}
\tilde{\phi}(x,y,{\overline y}) = \phi (x,y,{\overline y}) +
\beta (x,y,{\overline y}) + {\overline\beta }(x,y,{\overline y}) \;\; .
\end{equation}
The $\beta$ fields correspond to connections between group manifolds
analogously to $\phi$, which is a connection between space-times.
This should arise naturally when the underlying algebraic construction is
represented on Hilbert space.

A subtle but important distinction now arises with the model building
approach of Dubois-Violette et al.{\cite A}.
Strictly they are considering an extension
of the differential calculus to matrix algebras. However, we are considering
the set
of smooth functions over the group manifolds on each copy of
space-time. In this sense the matrix structure serves as a local basis against
which co-ordinates on the group manifolds are defined, i.e. on the tangent
space to the group. We do not seek to construct a derivation algebra on the
matrix algebra but rather on the co-ordinates defined locally by the
Lie algebra.

To implement the extension we begin at the level of the algebra and
generalize the differential operator $\delta$ to the set
\begin{equation}
\tilde{\delta} = (\delta ,\delta_{Q} ,\delta_{\overline Q} ) \;\; ,
\end{equation}
the subscripts $Q$ and $\overline Q$ introduced to signify the BRS and
anti-BRS operators. As we are restricting to classical groups all matrix
 elements will be real or complex. We are thus strictly dealing with
commutative Hopf algebras.
The set $\tilde{\delta}$ satisfy the Leibnitz rule where,
in order to satisfy nilpotency, the following are also implied
\begin{eqnarray}
\delta^{2} =\delta_{Q}^{2} = \delta_{\overline Q}^{2} = 0 \;\; ,
\hspace{1.5cm}\nonumber \\
{\rm and} \hspace{3cm}\;\; \nonumber \\
\{\delta ,\delta_{Q}\} = \{\delta ,\delta_{\overline Q}\} =
\{\delta_{Q}, \delta_{\overline Q}\} =0 \;\; .
\end{eqnarray}

\section{The BRS and anti-BRS Constraints}

Just as Baulieu et al.{\cite M} introduced ghosts as apriori geometrical fields
we are considering the dynamical fields as existing on an extended notion of
manifold described appropriately by a smooth algebra. We are thus
considering matrix elements
\begin{equation}
u_{ik}\in C^{\infty}(X\otimes G)\cong C^{\infty}(X)\otimes C^{\infty}(G)
\;\; ,
\end{equation}
on each copy of space-time. Furthermore, the groups are taken to be compact.
Thus we can be confident that our generalized connection will be a finite sum
of the form
\begin{equation}
\tilde{\omega} = \sum_{i} a^{i}\tilde{\delta} b^{i} =
\sum_{i} a^{i} (\delta + \delta_{Q} +\delta_{\overline Q} )b^{i} \;\; .
\end{equation}
This generalizes our notion of 1-form on the algebra where $a^{i}, b^{i}
\in {\cal A}_{t}$, ${\cal A}_{t}$ being the total algebra,
\begin{equation}
{\cal A}_{t}=C^{\infty}(X\otimes G_{1})\oplus C^{\infty}(X\otimes G_{2}) \;\; .
\end{equation}
Note that we do not assume that $G_{1}=G_{2}$.

To represent this on Hilbert space we must extend the representation of
Connes'{\cite{C,D,E}} onto the group manifolds. This requires constructing a
Clifford
algebra on the compact internal spaces as extensions of the space-time
Clifford algebra. Clearly there are parallels here with Kaluza-Klein
theory. For an $N$-dimensional internal manifold this will correspond
 to introducing a Clifford algebra belonging to the group $O(N)$. This need
not be associated with the underlying group upon which it is based. Compare
this with the construction of Baulieu et al.{\cite M} which follows the same
pattern.
We need not consider this to be a problem when it is recalled that the
Clifford algebra is introduced only as a means to represent differential
forms. In this sense the physical fields are not valued in this algebra.
Such a construction arises from demanding that the ghost fields carry both
gauge {\bf and} internal vector indicies.

Corresponding to the differential operator $\tilde{\delta}$ we write down
a Dirac operator
\begin{eqnarray}
\tilde{D}&=&D+Q+{\overline Q} \nonumber \\
&=&
\left (
\begin{array}{ll}
\,\;\; {\not\! \partial} \;\;\; \gamma_{5} M \\
\gamma_{5} M^{\ast} \;\; {\not\! \partial}
\end{array}
\right )
  +
\gamma_{5}\otimes
\left (
\begin{array}{cc}
{\not\! \partial}_{m} \;\; \tilde{\gamma_{5}}\xi \\
\tilde{\gamma_{5}} \xi^{\ast} \;\; {\not\! \partial}_{n}
\end{array}
\right )
  +
\gamma_{5}\otimes
\left (
\begin{array}{rr}
{\overline{\not\! \partial}}_{m} \;\; \tilde{\gamma_{5}}
{\overline \xi} \\
\tilde{\gamma_{5}} {\overline\xi^{\ast}} \;\;
{\overline{\not\! \partial}}_{n}
\end{array}
\right )  \;\; .
\label{DIRAC}
\end{eqnarray}
Here $\gamma_{5}$ is the usual chirality operator on four dimensional
space-time, $\tilde{\gamma}_{5}$ is the corresponding operator on the
group manifold, ${\not\!\partial}_{p},\; p=m,n$ are the operators on each
group manifold while $\xi$ is an $m\times n$ matrix yet to be specified.
Strictly, since $\tilde{\gamma}_{5}$ anti-commutes with the Clifford
algebras on both group manifolds we should write $\tilde{\gamma}_{5}
=\gamma_{5}^{m}\otimes\gamma_{5}^{n}$, which will be understood
from now on. For clarity we can explicitly set out the K-cycles of the
model. Writing the total algebra (28) as
\begin{equation}
C^{\infty}(X)\otimes (C^{\infty}(G_{1})\oplus C^{\infty} (G_{2}))
\end{equation}
the generalization of our notion of internal space becomes clear.
The Dirac operator on the continuous manifolds takes the form
${\not\! \partial} + \gamma_{5} ({\not\! \partial}_{p} +
{\overline{{\not\! \partial}}}_{p} )$, operating on the space of square
integrable functions on space-time and smooth
functions on the group manifolds. The functions on the group manifolds
have the form $u_{ik}(y)$ corresponding to matrix elements in the
space of complex matrices $M_{p}({\bf C})$. On the
discrete space we introduce the Dirac operator
\begin{equation}
\left (
\begin{array}{ll}
\; 0 \;\;\; M \\
 M^{\ast} \; 0
\end{array}
\right )
  +
\tilde{\gamma}_{5} \{
\left (
\begin{array}{cc}
0 \;\; \xi \\
 \xi^{\ast} \;\; 0
\end{array}
\right )
  +
\left (
\begin{array}{rr}
0 \;\;
 {\overline \xi} \\
{\overline \xi}^{\ast} \;\;
0
\end{array}
\right ) \} \;\; .
\end{equation}
acting on the Hilbert space ${\bf C}^{m}\oplus {\bf C}^{n}$. The
extension of the internal Dirac operator introduces the notion of a
connection between group manifolds analogously to that between
space-times. The K-cycle for both continuous and discrete contributions
is now described by the operator (\ref{DIRAC}). (Note that the ``product"
K-cycle
is actually between the classical K-cycle for $C^{\infty}(X)$ and our
generalized concept of internal space). It is important to observe that
since $O(N)$ has fundamental group $Z_{2}$ for all $N > 2$ there will
be no
problem in defining the notion of spinors on the continuous internal
manifolds for non-abelian groups.

Representing the connection 1-form on Hilbert space we have
\begin{equation}
\rho (a^{i})=
\left (
\begin{array}{cc}
A_{0}^{i} \;\; 0 \\
0 \;\; B_{0}^{i}
\end{array}
\right ) \;\; , \;\;
\rho (b^{i})=
\left (
\begin{array}{cc}
A_{1}^{i} \;\; o \\
0 \;\; B_{1}^{i}
\end{array}
\right )
\end{equation}
so that
\begin{eqnarray}
\pi (\tilde{\omega} )&=& \sum_{i} \rho (a^{i})[D+Q+{\overline Q},
\rho (b^{i})]
\nonumber \\
&=&
\sum_{i}\left \{
\left (
\begin{array}{ll}
A_{0}^{i}{\not\!\partial}A_{1}^{i} \;\;
\gamma_{5} A_{0}^{i}(MB_{1}^{i}-A_{1}^{i}M) \\
\gamma_{5}B_{0}^{i}(M^{\ast}A_{1}^{i}-B_{1}^{i}M^{\ast}) \;\;
B_{0}^{i}{\not\!\partial}B_{1}^{i}
\end{array}
\right ) + \gamma_{5}\otimes
\left (
\begin{array}{cc}
A_{0}^{i}{\not\!\partial}_{m} A^{i}_{1} \;\;
\tilde{\gamma_{5}}A_{0}^{i}(\xi B_{1}^{i}-A_{1}^{i}\xi ) \\
\tilde{\gamma_{5}}B_{0}^{i}(\xi^{\ast}A_{a}^{i}-B_{1}^{i}\xi^{\ast} ) \;\;
B_{0}^{i}{\not\!\partial}_{n} B_{1}^{i}
\end{array}
\right )
\right . \nonumber \\ &&\hspace{5cm}+ \left .  \gamma_{5}\otimes
\left (
\begin{array}{cc}
A_{0}^{i}{\overline{\not\!\partial}}_{m} A^{i}_{1} \;\;
\tilde{\gamma_{5}}A_{0}^{i}({\overline\xi} B_{1}^{i}-A_{1}^{i}
{\overline\xi} ) \\
\tilde{\gamma_{5}}B_{0}^{i}(
{\overline\xi}^{\ast}A_{a}^{i}-B_{1}^{i}{\overline\xi}^{\ast} ) \;\;
B_{0}^{i}{\overline{\not\!\partial}}_{n} B_{1}^{i}
\end{array}
\right )
\right \} \nonumber \\
&=&
\left (
\begin{array}{ll}
A \;\; \gamma_{5}\phi \\
\gamma_{5}\phi^{\ast} \;\; B
\end{array}
\right ) +\gamma_{5}\otimes
\left (
\begin{array}{cc}
C_{A} \;\; \tilde{\gamma}_{5}\beta \\
\tilde{\gamma}_{5}\beta^{\ast} \;\; C_{B}
\end{array}
\right ) + \gamma_{5}\otimes
\left (
\begin{array}{rr}
{\overline C}_{A} \;\; \tilde{\gamma}_{5}{\overline\beta} \\
\tilde{\gamma}_{5}{\overline\beta}^{\ast} \;\; {\overline C_{B}}
\end{array}
\right ) \;\; .
\end{eqnarray}
We thus have ghosts and anti-ghosts for each gauge group ($C_{A,B},
{\overline C}_{A,B}$), gauge fields ($A,B$) and now matter fields, $\phi$,
with corresponding extensions to ghost and anti-ghost matter fields
($\beta, {\overline\beta}$) consistent with the interpretation of
$\phi$ as an extended notion of gauge field. Note that the hermiticity
requirement on $\pi (\tilde{\omega} )$ connects us with the Lie
algebra for each group.

Proceeding as with Baulieu et al.{\cite M} we will now consider the generalized
curvature, imposing the Cartan-Maurer condition to derive the constraints.
Using the conditions (25) the complexity of this can be greatly reduced.
We have the curvature at the level of the algebra
\begin{equation}
\Theta =\tilde{\delta}\tilde{\omega} + \tilde{\omega}^{2} \;\; ,
\end{equation}
where
\begin{eqnarray}
\tilde{\delta}\tilde{\omega} &=& (\delta +\delta_{Q} +
\delta_{\overline Q} )\sum_{i} (a^{i}(\delta + \delta_{Q} +
\delta_{\overline Q} )b^{i} ) \nonumber \\
&=& \delta a^{i}\delta b^{i} + \delta a^{i}\delta_{Q} b^{i} +
\delta a^{i} \delta_{\overline Q} b^{i} + \delta_{Q} a^{i}\delta b^{i}
\nonumber \\
&& +\delta_{Q} a^{i}\delta_{Q} b^{i} + \delta_{Q} a^{i}
\delta_{\overline Q} b^{i} + \delta_{\overline Q} a^{i}\delta b^{i} +
\delta_{\overline Q} a^{i}\delta_{Q} b^{i} \nonumber \\
&& + \delta_{\overline Q} a^{i}\delta_{\overline Q} b^{i}
\end{eqnarray}
terms of the form $a^{i}\delta\delta_{Q} b^{i}$ vanishing. This
illustrates the utility of working on the algebra where the calculus
is well defined. Using the nomenclature of Baulieu et al.{\cite M} we set to
zero
the terms of $\pi (\Theta )$ proportional to $dx^{\mu}\wedge dy^{p}$,
$dx^{\mu}\wedge d{\overline y}^{p}$, $dy^{p}\wedge dy^{p^{\prime}},
dy^{p}\wedge d{\overline y}^{p^{\prime}}$ and
$d{\overline y}^{p}\wedge  d{\overline y}^{p^{\prime}}$ ,
noting that $y^{p}$ are
co-ordinates on one of the two group manifolds. (Note, however, that
we do not get mixed forms between groups, e.g. $dy^{p}_{m}\wedge
dy^{p^{\prime}}_{n}$). Since matter fields are also present we will get
forms proportional to $dy^{p}$ and $d{\overline y}^{p}$,
corresponding to the covariant derivatives of the matter fields on the
group manifold. These are also set to zero.  We recall that such forms
are still 2-forms in the generalized sense of Connes'
construction{\cite{C,D,E}}.
Setting
\begin{eqnarray}
S_{A} &=& \gamma_{m}^{p}\partial_{p}\;\; (\leftrightarrow dy^{p}_{m}
\partial_{p} ) \nonumber \\
S_{B} &=& \gamma_{n}^{p}\partial_{p}\;\; (\leftrightarrow dy^{p}_{n}
\partial_{p} ) \;\; ,
\end{eqnarray}
we derive the constraints:
\begin{eqnarray}
&S_{A}A^{\mu}=D^{\mu}C_{A} \;\;\; &S_{B}B^{\mu}=D^{\mu}C_{B}
\nonumber \\
&S_{A}\phi= -C_{A}(\phi +M) &S_{B}\phi =(\phi +M)C_{B}
\nonumber \\
&S_{A}\phi^{\ast} =(\phi^{\ast} +M^{\ast})C_{A}
&S_{B}\phi^{\ast} = -C_{B}(\phi^{\ast} + m^{\ast}) \nonumber \\
&S_{A}\beta =-C_{A}(\beta +\xi ) & S_{B}\beta = (\beta + \xi )C_{B}
\nonumber \\
&S_{A}\beta^{\ast} =(\beta^{\ast} +\xi^{\ast} )C_{A}
&S_{B}\beta^{\ast} =-C_{B}(\beta^{\ast} +\xi^{\ast} ) \nonumber \\
&S_{A}{\overline \beta}=-C_{A}({\overline\beta}+{\overline\xi})
&S_{B}{\overline\beta}=({\overline\beta}+{\overline\xi})C_{B} \nonumber \\
&S_{A}{\overline\beta}^{\ast} =({\overline\beta}^{\ast}
+{\overline\xi}^{\ast} )C_{A}
&S_{B}{\overline\beta}^{\ast}=-C_{B}({\overline\beta}^{\ast}
+{\overline\xi}^{\ast} ) \nonumber \\
&S_{A}C_{A}=-{1/2}[C_{A},C_{A}]  &S_{B}C_{B}=-{1/2}[C_{B},C_{B}]
\end{eqnarray}
and similarly for ${\overline S}_{A}$ and ${\overline S}_{B}$,
for example
\begin{eqnarray}
&{\overline S}_{A}A^{\mu}=D^{\mu}{\overline C}_{A} \;\;\;
&{\overline S}_{B}B^{\mu}=D^{\mu}{\overline C}_{B} \nonumber \\
&{\overline S}_{A}\beta =-{\overline C}_{A}(\beta +\xi )
&{\overline S}_{B}\beta =(\beta +\xi ){\overline C}_{B} \nonumber \\
&{\overline S}_{A}{\overline C}_{A}=-{1/2}[{\overline C}_{A},
{\overline C}_{A}]
&{\overline S}_{B}{\overline C}_{B}=-{1/2}[{\overline C}_{B},
{\overline C}_{B}] \;\; {\rm etc}
\end{eqnarray}
and the cross term
\begin{equation}
S_{X}{\overline C}_{X}+{\overline S}_{X}C_{X} +
[{\overline C}_{X}, C_{X} ] =0 \;\; ,
\end{equation}
where $X= A,\; B$. Note that we are implicitly working on the correct space
of forms not having considered the differential ideal $J$. That is, we have
not exhibited the auxiliary terms which arise in the curvature due to
the ambiguity in the representation $\pi${\cite D}. These auxiliary terms are
crucial for the correct determination of the Higgs potential but do
not impact on the constraint equations.

As expected we reproduce the usual BRS and anti-BRS constraints on each
gauge group. However, we now also have constraints involving
the new field $\beta$. The element which is missing is the Zinn-Justin
auxiliary field required to define $S_{X}{\overline C}_{X}$, which
is left arbitrary by the constraints. Baulieu et al.{\cite M} introduce this
field, $b$, so that
\begin{equation}
S{\overline C} =b \;\; ,
\end{equation}
and so close the algebra. Rather than introduce a new field we shall propose
that the role of $b$ is taken instead by $\beta$. This appears to be
consistent as $\beta$ should not appear in the physical equations of
motion and thus must be treated as auxiliary. Furthermore, we now have
a means by which the field $b$, defined in the adjoint of a given
gauge group, may be decomposed into constituent parts. This is a
natural consequence of dealing with representations on Hilbert space
where fermionic elements are the basic building blocks.

To recover the Zinn-Justin auxiliary field we apply the constraints (37)
so that ( passing to the symmetric phase $\tilde{\beta} =\beta +\xi$)
\begin{eqnarray}
\left .
\begin{array}{c}
S_{A}\tilde{\beta} =-C_{A}\tilde{\beta} \; \\
S_{A}\tilde{\beta}^{\ast} =\tilde{\beta}^{\ast} C_{A} \;
\end{array}
\right \} \Rightarrow
S_{A}(\tilde{\beta}\tilde{\beta}^{\ast} ) =
(S_{A}\tilde{\beta} )\tilde{\beta}^{\ast} +
\tilde{\beta} S_{A}\tilde{\beta}^{\ast}
 = [\tilde{\beta}\tilde{\beta}^{\ast} ,C_{A}]
\end{eqnarray}
so that in this case $b=\tilde{\beta}\tilde{\beta}^{\ast}$. Similarly,
\begin{eqnarray}
\left .
\begin{array}{c}
S_{B}\tilde{\beta} =\tilde{\beta} C_{B} \; \\
S_{B}\tilde{\beta}^{\ast} =-C_{B}\tilde{\beta}^{\ast} \;
\end{array}
\right \} \Rightarrow
S_{B}(\tilde{\beta}^{\ast} \tilde{\beta}) =
(S_{B}\tilde{\beta}^{\ast} )\tilde{\beta} +
\tilde{\beta}^{\ast} S_{B}\tilde{\beta}
 = [\tilde{\beta}^{\ast}\tilde{\beta} ,C_{B} ]
\end{eqnarray}
so for the $B$ gauge field  sector we have the identification
$b=\tilde{\beta}^{\ast}\tilde{\beta}$. ( An alternative is to identify
the $A$ and $B$ gauge fields. The $\beta$ fields are then necessarily in
self-adjoint $m\times m$ representations of the gauge group{\cite F}. It
follows
then from the constraints (37) that
\begin{equation}
S\tilde{\beta} ={1/2}[\tilde{\beta} ,C] \;\; .
\end{equation}
The factor of $1/2$ is spurious since we should strictly make this
identification at
the level of the connection, not the constraints.
This approach, however, is not a preferred option as it can be shown that then
\begin{equation}
\pi (\delta ({\rm ker}\pi )^{1})=0
\end{equation}
on the internal space
and as a result one will no longer yield a Higgs potential in a
symmetry breaking form{\cite R}). In our construction we also have the ``charge
conjugate'' field $\overline\beta$. We can, as above, construct the
conjugate auxiliary field. The most general such term which we can construct
has the
form
$b=
(\tilde{\beta}+\tilde{\overline{\beta}})(\tilde{\beta}+\tilde{\overline{\beta}})^{\ast}
+(\tilde{\beta}+\tilde{\overline{\beta}})^{\ast}(\tilde{\beta}+
\tilde{\overline{\beta}})$.
It follows from the constraints that
\begin{eqnarray}
S_{X}(b) &=& [b, C_{X}] \nonumber \\
{\overline S}_{X}(b) &=& [b,{\overline C}_{X}] \;\; .
\end{eqnarray}
We now make
the identification
\begin{eqnarray}
S_{X}{\overline C}_{X} &=& b \nonumber \\
{\overline S}_{X}C_{X} &=& -b-[{\overline C}_{X},C_{X}] \;\; .
\label {star}
\end{eqnarray}
Note that it is not necessary to keep track of which gauge sector we are
considering since, for example
\begin{eqnarray}
S_{A}(\tilde{\beta}^{\ast}\tilde{\beta} ) &=&
(S_{A}\tilde{\beta}^{\ast} )\tilde{\beta} +
\tilde{\beta}^{\ast} S_{A}\tilde{\beta} \nonumber \\
& =&\tilde{\beta}^{\ast} C_{A}\tilde{\beta} -
\tilde{\beta}^{\ast} C_{A}\tilde{\beta} \nonumber \\
& =& 0 \;\; .
\end{eqnarray}
 Applying the nilpotency conditions (25),
which on the correct space of forms are represented as
\begin{eqnarray}
S_{X}^{2}={\overline S}_{X}^{2} =0 \hspace{0.8cm}\nonumber \\
S_{X}{\overline S}_{X} + {\overline S}_{X} S_{X} =0 \;\; ,
\end{eqnarray}
it follows that
\begin{eqnarray}
S_{X}(b) &=& 0 \nonumber \\
{\overline S}_{X}(b)&=&[b,{\overline C}_{X} ] \;\; .
\end{eqnarray}
so that we have constrained $[b,C_{X}]=0$ by this choice.
We can thus recover in a very natural way the set of BRS and anti-BRS
equations on each gauge group where now the matter field constraints
on $\phi$ arise at the same level as those of the gauge fields. The
Zinn-Justin auxiliary scalar now appears as a result of this fundamental
Higgs-gauge field unification rather than as an additional field. To
insure the auxiliary nature of $\beta$ we also impose the constraint
\begin{equation}
D_{\mu}{\beta} =0 \;\; .
\end{equation}
That $\beta$ should be trivial on space-time is consistent with the
notion that it is a connection between group manifolds only.
(It is tempting to try a more symmetric identification than
({\ref {star}}) and extend the choice of Baulieu et al.{\cite M} such that
\begin{equation}
S_{X}b=0 \;\; , \;\; {\overline S}_{X}= [b,{\overline C}_{X}]
\end{equation}
 now includes
\begin{equation}
S_{X}b^{c}=[b^{c}, C_{X}] \;\; , \;\; {\overline S}_{X}b^{c}=0 \;\; .
\end{equation}
where $b=\tilde{\beta}^{\ast}\tilde{\beta} +\tilde{\beta}\tilde{\beta}^{\ast}$
and $b^{c}=\tilde{\overline{\beta}}^{\ast}
\tilde{\overline{\beta}} +\tilde{\overline{\beta}}
\tilde{\overline{\beta}}^{\ast}$.
This would extend the construction of Baulieu et al.{\cite M}
to include $b^{c}$ in such
a way that $b^{c}$ is a constant of the motion in the anti-ghost sector in
a complimentary way to $b$ on the ghost sector. However, it is not
possible to make this identification and keep compatability with the
remaining algebra, for example, requiring that $S_{X}^{2}{\overline C}_{X}
=0$. For this reason the choice of ({\ref {star}}) is the appropriate one).

While this is sufficient when restricted to the gauge fields
the existence of Higgs scalars connecting different gauge fields requires
independent consideration. This is because generalized 2-forms in the
construction of Connes'{\cite {C,D,E}}
will now arise which are scalars in the traditional
sense. This is just the origin of the Higgs potential in Connes model
building prescription. However, we now also have $\beta$ and
$\overline\beta$ terms which will contribute. We thus expect to
generalize our potential which will take the form
\begin{equation}
V(\phi ,\beta ,{\overline\beta})=V(\phi )+V(\beta ,{\overline\beta})
 +{\rm\bf mixing \;\; terms} \;\; .
\end{equation}
The mixing terms demand that $\beta$ and $\overline\beta$ remain the
fundamental auxiliary fields, rather than $b$. Elimination
of $\beta$ and $\overline\beta$ will now have a dramatic effect,
introducing additional interaction terms between the Higgs, ghost
and gauge fields.
The actual form taken by the interaction terms
will be very much model dependent and so will be investigated
in specific examples currently in preparation. The important consequence of
this is that
we have extended the notion of gauge fixing into the Higgs sector, consistent
with gauge-Higgs field unification.

\section{The Quantum Lagrangian}

We wish now to consider the most general allowable BRS and
anti-BRS invariant lagrangian from which the BRS/anti-BRS admissable
gauges may be considered. Again, restricting ourselves at first to the
underlying algebra will greatly simplify this construction. We know that
the form of our Higgs potential before elimination of the auxiliary terms
derives from the square norm of the curvature. In looking at the quantum
lagrangian we wish to avoid the introduction of terms like $\phi^{2}$ with
arbitrary
coefficients which are not forbidden by BRS/anti-BRS invariance but which could
spoil the Higgs potential. We therefore choose to
maintain the action as a functional of the curvature{\cite A}.
The
simplest such term which can be constructed, which is $\delta_{Q}$ and
$\delta_{\overline Q}$ invariant and of dimension 4 is
\begin{equation}
\delta_{Q}\delta_{\overline Q}\Theta =
\delta_{Q}\delta_{\overline Q} (\tilde{\delta}\tilde{\omega} +
\tilde{\omega}^{2} ) \;\; .
\label {star2}
\end{equation}
To be physical we require the ghost number to be zero, insuring that
we are in the correct cohomology class. Assigning a ghost number of
$1$, say, to $\delta_{Q}$ and $-1$ to $\delta_{\overline Q}$ we see
that this restriction reduces ({\ref {star2}}) to
\begin{equation}
\delta_{Q}\delta_{\overline Q} (\delta\omega +\omega^{2}
+ \alpha\sum_{ij} (a^{i}\delta_{Q} b^{i}a^{j}
\delta_{\overline Q} b^{j} + a^{i}\delta_{\overline Q}
b^{i}a^{j}\delta b^{j} ))
\label{noghost}
\end{equation}
where $\omega =\sum_{i} a^{i}\delta b^{i}$ is the usual, classical,
connection used in Connes' construction{\cite E}. Here $\alpha$ may be
introduced as
an arbitrary constant, the choice of which corresponds to the gauge
choice{\cite M}.
Note that these requirements on ghost number and $\delta_{Q}$ and
$\delta_{\overline Q}$ invariance are imposed at the, well behaved, level of
the
algebra. Representing this on Hilbert space will, however, introduce terms with
non-zero ghost number and which are not invariant under $S_{X}$ and ${\overline
S}_{X}$. This results from treating BRS/anti-BRS invariance in a unified way on
the
algebra but then separating contributions on the two gauge groups
when physical fields are constructed. That the physical set of operators is a
restricted set of those on the differential algebra is a recurring theme in
all applications of noncommutative geometry.

Representing this on Hilbert space now requires some care. The reason is that
contributions such as $\sum_{i} a^{i}\delta\delta_{Q} b^{i}$
need not vanish due to a
lack of cross terms. To represent such terms the representation $\pi$
must be extended to accomodate our generalized Leibnitz rule. We will thus
define
\begin{equation}
\pi (\delta_{Q}\delta a^{i} ) = \{ Q, [D, \rho (a^{i})]\} \;\;\;\;
\pi (\delta_{\overline Q}\delta a^{i}) = \{ {\overline Q},[D,\rho( a^{i})]\}
\nonumber
\end{equation}
and
\begin{equation}
\pi (\delta_{Q}\delta_{\overline Q}\delta a^{i} ) = [Q,\{ {\overline
Q},[D,\rho (a^{i})]\} ] \;\; ,
\end{equation}
which now encodes correctly the Leibnitz rule on each copy of space-time and
on the matrix derivative connecting space-times.

To extract the physical fields from (\ref{noghost}) we note that the action
in non-commutative geometry which is given by the Dixmier trace can be
written equivalently as{\cite F}
\begin{equation}
I={1/8}\int d^{4}x{\rm Tr}({\rm tr}(\pi^{2} (\theta ))) \;\; ,
\label{action}
\end{equation}
$\theta$ being the usual curvature in Connes' construction ($\theta
=\delta\omega + \omega^{2}$), Tr is taken over the matrix structure and tr is
taken over the Clifford algebra. We extend this now to include the ``quantum
term"
\begin{equation}
I_{\rm quantum} ={1/8}\int d^{4}x{\rm Tr}({\rm tr}[\pi^{2} (\theta^{\prime} )
+\pi (\delta_{Q}\delta_{\overline Q}\Theta )_{\rm zero \; ghost \; no.} ])
\;\; ,
\label {ql}
\end{equation}
where we write $\theta^{\prime}$ since we must also now accomodate the extended
potential (53).
We need thus consider only those terms which survive under the two trace
operations. Note, however, that we will take the same liberty as
Baulieu et al.{\cite M}
and retain the fields $C_{X}$ and ${\overline C}_{X}$ as well as $S_{X}$ and
${\overline S}_{X}$ as differential forms. To this extent we only take tr over
the four dimensional space-time Clifford algebra. In terms of Dixmier
traces we see then that the action need only be defined with respect to
the classical K-cycle of continuous functions on space time tensored
with a discrete internal space. There is, therefore, no problem with
$(d,\infty )$ summability, consistent with imposing the Cartan-Maurer
condition.

The simplest expression to calculate is $\pi (\delta_{Q}\delta_{\overline
Q}\delta\omega )$ for which we find (retaining the $S_{X}$ and ${\overline
S}_{X}$
invariant terms)
\begin{equation}
S_{A}{\overline S}_{A}(M\phi^{\ast} + \phi
M^{\ast}) +
S_{B}{\overline S}_{B}(M^{\ast}\phi + \phi^{\ast} M) \;\;,
\end{equation}
where since we require global gauge invariance terms of the form
$S_{A}{\overline S}_{A}(\partial_{\mu} A^{\mu})$ have been excluded.
Similarly for $\pi (\delta_{Q}\delta_{\overline Q}\omega^{2} )$ the result is
\begin{equation}
S_{A}{\overline S}_{A}(A_{\mu}^{2} + \phi\phi^{\ast}) +
S_{B}{\overline S}_{B}(B_{\mu}^{2} + \phi^{\ast}\phi ) \;\; ,
\label {phi}
\end{equation}
where we note again that since we require global gauge invariance
 $\phi\phi^{\ast}$ and
$\phi^{\ast}\phi$ must transform as singlets. Consequently
$S_{A}{\overline S}_{A}(\phi\phi^{\ast})=
S_{B}{\overline S}_{B}(\phi^{\ast}\phi )=0$.
Greater care must be taken in determining the
final term to avoid contributions with non-zero ghost number. Retaining the
$S_{X}$
and ${\overline S}_{X}$ invariant terms we find,
after lengthy algebra, the relevant contributions for
$\pi (\delta_{Q}\delta_{\overline Q} [\sum_{ij} (a^{i}\delta_{Q}
b^{i}a^{j}\delta_{\overline Q}b^{j} +a^{i}\delta_{\overline
Q}b^{i}a^{j}\delta_{Q}b^{j})])$:
\begin{equation}
S_{A}{\overline S}_{A}(C_{A}{\overline C}_{A} + {\overline C}_{A}C_{A}
+\beta{\overline \beta}^{\ast} + {\overline \beta}\beta^{\ast} ) +
S_{B}{\overline S}_{B}(C_{B}{\overline C}_{B} + {\overline C}_{B}C_{B}
+\beta^{\ast}{\overline \beta} + {\overline \beta}^{\ast}\beta ) \;\; ,
\end{equation}
where we eliminate the $\beta$ terms analogously to the $\phi$'s in
(\ref{phi}). The removal of these terms is consitent with hermiticity
of the ``quantum term" as implied by (46). In
all these expressions the auxiliary fields associated with the representation
$\pi$
have been surpressed. Being at fourth order, such auxiliary contributions are
tediously complex so there is little utility in expressing their general form.
We see  then that we are left with only the most obviously
$S_{X}$ and ${\overline S}_{X}$
invariant terms as physical contributions. Clearly, a projection onto the
correct space of forms will involve greater mathematical rigour and
complexity. Nevertheless, we would expect the general form of the above
expressions to be maintained since the requirement of global gauge invariance
has removed terms dependent on $\phi$ and $\beta$ for which we would expect
nontrivial contributions.

We can thus recover, in a very natural way, the quantum term of Baulieu et
al.{\cite M} in the case of symmetric gauges.
Rather than an exhaustive construction of all possible allowable polynomials
of the fields we can develop the same result directly from our generalized
curvature, consistently including the contribution from the Higgs sector.

\section{Application to Anomalies}

The constructions of the previous sections demonstrate that significant
simplifications in the construction of field theoretic models arise by first
restricting considerations to the underlying universal differential algebra.
This highlights the quantity of relevant information contained at this level of
the model building prescription. Only when a particular representation, $\pi$,
is chosen do we encounter complications. This is a reflection of the wide and
as
yet non-unique choice of representation on the Hilbert space of spinors.

Using this observation we can direct our attention to the important topic of
anomalies. Since anomaly cancellations are dependent on the choice of fermionic
representations and not fermion masses we see that we require only that
information which is entered at the level of the algebra ${\cal A}_{t}$.
Strictly, we should be dealing with a local BRS operator but this extension is
not considered to be problematic.

It is known that consistency conditions for anomalies appear as cohomological
equations for the BRS operator ( we will neglect the anti-BRS operator for
simplicity). In this regard we will utilize the trivial cohomological
structures
on the universal differential algebra to simplify considerations for otherwise
complex models involving multiple gauge fields, Higgs scalars etc. We will
follow
the prescription of Dubois-Violette, Talon and Viallet{\cite {DTV}} for
constructing
solutions of the Wess-Zumino consistency conditions{\cite {WZ}}.
We consider the free graded
commutative algebra generated by:
\begin{eqnarray}
A &=& \sum_{i} a^{i}\delta b^{i} \nonumber \\
F &=& \delta A + A^{2} \nonumber \\
\chi &=& \sum_{i} a^{i}\delta_{Q} b^{i} \nonumber \\
\phi &=& \delta\chi
\end{eqnarray}
which we denote as $\cal C$. $A$ and $\chi$ comprise the generalized
potential
\begin{equation}
\tilde{A} = \sum_{i} a^{i}(\delta+\delta_{Q} )b^{i} = A + \chi \;\; .
\end{equation}
The generalized curvature is then
\begin{equation}
\tilde{F} = (\delta +\delta_{Q} )\tilde{A} + \tilde{A}^{2} \;\; .
\end{equation}
Imposing the Maurer-Cartan condition at the level of the universal
differential algebra it follows that
\begin{equation}
\delta_{Q}\chi = -\chi^{2} \;\; , \;\;
\delta_{Q} A = -\phi-A\chi-\chi A \;\; ,
\label{terms}
\end{equation}
and therefore $ \delta_{Q} F = [F, \chi ]$. Note that as before, when
considering the quantum lagrangian, this BRS algebra is {\underline
{not}} equivalent to that derived on the physical Hilbert space. This
follows since the above relations (\ref{terms}) imply that those terms
dependent on $\beta$ will not appear in the generalized potential
(53). From this we see that the auxiliary nature of $\beta$ has been
made manifest. Consistent with Connes' model building prescription
$\beta$ remains as part of the generalized potential due only to the
nature of the representation on Hilbert space. In this context, the
removal of $\beta$ via the equations of motion is motivated by purely
algebraic considerations and not only phenomenological consistency.
This Zinn-Justin scalar has thus been reduced to one of several
auxiliary terms inherent in the model building scheme.

Using $A,\delta A,\chi,\delta\chi$ as a free system of generators of
$\cal C$ it follows that the algebra $({\cal C},\delta)$ is a
contractible differential algebra. Similarly, $A,\chi , (\delta +
\delta_{Q} )A, (\delta + \delta_{Q} )\chi$ is a free system of
generators of $\cal C$ so $({\cal C}, \delta +\delta_{Q} )$ is
contractible. As can be seen from the BRS relations (\ref{terms}) the
cohomology of $\delta_{Q}$ is not trivial. It follows that $({\cal C},
\delta_{Q} )$ is the skew tensor product of the contractible algebra
$A, \delta_{Q} A$ and the algebra $\chi , F$. The $\delta_{Q}$
cohomology thus reduces to that on $\chi$ anf $F$. From this it can be
shown that the $\delta_{Q}$ cohomology reduces to sets of invariant
polynomials in these fields. Thus anomalies and Schwinger terms are
obtained from such invariants{\cite {DTV}}.

This type of analysis remains faithful for simple
gauge theories. But we now
have the additional step of identifying with complex models involving
multiple gauge fields and Higgs scalars via the representation $\pi$.
On the physical space such a simple cohomological treatment need not
hold. However,
as we saw in the previous section, representations of appropriate
polynomial functions constructed on the universal differential
algebra compactly describe all relevant polynomial contributions,
including scalar contributions, on the physical space. There was no
need to exhaustively explore all possible polynomial functions.
Similarly, representing invariant polynomial functions pertaining to
anomalies, which are simply described on the universal differential
algebra,
will encode all the relevant information on these
contributions on the physical space
without the need for an arbitrary exhaustive search; the
relevant physical terms being extracted as before by such requirements
as $S_{X}$ or global gauge invariance. This provides an important tool
for probing complicated models. We would still expect model dependency
in this approach as the $\delta_{Q}$ cohomology is dependent on the
number of $U(1)$ factors in the model in question.

\section{Some Comments on Auxiliary Terms}

The astute reader will observe that the gauge terms appearing in (\ref{phi})
are
normally considered as auxiliary in Connes' construction{\cite E}
but are retained in the
``quantum" sector of our action (\ref{ql}). This is not inconsistent when we
examine the nature of these auxiliary terms more closely.

That problems arise from such contributions can be seen if one considers the
space
of generalized 2-forms in the classical case{\cite G}
\begin{equation}
{\hat{\Omega}}^{2}
({\cal A}_{t})=[{\hat{\Omega}}^{2}({\cal F})\otimes \rho ({\cal A}) +
{\cal F}\otimes {\hat{\Omega}}^{2}({\cal A})] \oplus
{\hat{\Omega}}^{1}({\cal F})\otimes {\hat{\Omega}}^{1}({\cal A}) \;\; ,
\label{o2}
\end{equation}
where we recall that ${\cal A}_{t} = {\cal F}\otimes {\cal A}$, corresponding
to
the product geometry of smooth functions on space-time and a discrete internal
space. The first term gives the usual gauge field curvature tensor, the second
corresponding to, the square root of, the Higgs potential and the last
responsible
for the covariant derivative on the mater fields. The addition in brackets is
not
direct because space-time 0-forms and 2-forms mix. Consequently, in the tensor
product geometry, the Higgs potential contributions must be disentangled from
the
0-forms of the gauge sector. This is the fundamental reason for the need of at
least two fermionic families to insure non-trivial projections onto the correct
orthocomplement. We can write down these 0-form contributions explicitly in,
say,
the ``$A$" gauge field sector, which are{\cite F}:
\begin{equation}
\sum_{i} A_{0}^{i}{\not\!\partial}^{2}A_{1}^{i} + \partial^{\mu} A_{\mu} +
A_{\mu}^{2} \;\; .
\end{equation}
The first appears when one calculates $\pi (\sum_{i}\delta a^{i}\delta b^{i} )$
as
an element which cannot be expressed in terms of the physical fields while the
remaining two arise from the Clifford algebra, e.g.
\begin{equation}
{\not\!\partial}\gamma^{\nu} A_{\nu} = {1/2}\sigma^{\mu\nu}(\partial_{\mu}
A_{\nu}
- \partial_{\nu} A_{\mu} ) -\partial^{\mu} A_{\mu} \;\; ,
\end{equation}
where $\{ \gamma^{\mu} , \gamma^{\nu}\} = -2\delta^{\mu\nu}$ in four
dimensional
Euclidean space. Thus $\partial^{\mu} A_{\mu}$ is associated with $\pi
(\delta\rho
)$ while similarly $A_{\mu}^{2}$ derives from $\pi (\rho^{2} )$. The important
point, however, is that since $\sum_{i} A_{0}^{i}{\not\!\partial}^{2}A_{1}^{i}$
is
an arbitrary function the terms of the scalar Higgs potential could be absorbed
into it. Thus it is this term which is responsible for the need  for careful
consideration on the correct space of forms. The remaining terms are,
nevertheless, unsavoury, implying that space-time 2-forms in Connes
construction
will not correspond to our usual notion of 2-form unless eliminated.

This brings us now to the ``quantum term"
$\pi (\delta_{Q}\delta_{\overline Q}\Theta )_{\rm zero \; ghost \; no.}$ of our
action. Unlike in (\ref{action}) this term is not quadratic in the curvature.
Immediately this implies that terms such as $A_{\mu}^{2}$ will play a more
fundamental role, being Lorentz scalars. It seems natural then to treat
$\pi (\delta_{Q}\delta_{\overline Q}\Theta )_{\rm zero \; ghost \; no.}$
independently to the classical action, a point emphasised by it being dependent
on
$S_{X}$ and ${\overline S}_{X}$, unlike the classical case. We thus comfront
the
interesting point of treating as auxiliary the fields
$\partial^{\mu} A_{\mu} +A_{\mu}^{2}$ whose elements are well defined in terms
of
known fields and do not otherwise adversely affect the action. In our context
they
find a role which need not violate our usual notion of differential forms when
treated independently from them. This notion is consistent with the imposition
of
the Maurer-Cartan form in deriving the BRS/anti-BRS constraints which can be
considered as a ``horizontality condition". The ``quantum" contribution then
comes
from the vertical, orthogonal, sector generated by non-vanishing terms in the
BRS/anti-BRS generators. An interesting implication of this is that the
space-time Dirac operator
in the vertical sector will no longer be nilpotent, implying that higher
order terms will be unphysical.

Turning to the BRS and anti-BRS region we would expect generalized 2-forms to
encounter the same problems as the Dirac operator, $D$, as expressed in
(\ref{o2}). Since $\beta$ type terms will contribute to the scalar potential
these, as with the $\phi$ scalars, will necessarily carry information
pertaining
to several fermionic families. Consistency demands that these family mixing
matrices be identical for $\phi$ and $\beta$. In addition to this we must also
account for the nilpotency conditions encoded in (25) between the operators
$\delta$, $\delta_{Q}$ and $\delta_{\overline Q}$. That is, there will exist
forms for wich, for example,
\begin{equation}
\pi (\delta_{Q}\delta_{\overline Q} a^{i} + \delta_{\overline Q}\delta_{Q}
a^{i})\not= 0
\;\; .
\end{equation}
This simply tells us that auxiliary terms will arise in accordance with mixing
terms, as expressed in the potential (53). (Note that, by the definition of our
generalized Dirac operator $\tilde D$ (31), there will be no such problem
between
$\left (
\begin{array}{cc}
{\not\!\partial} \;\;\;\;\;\; \\
\;\;\;\;\;\; {\not\!\partial}
\end{array}
\right )$
and $Q$ or ${\overline Q}$ so that in this case only the connection between
space-times and group manifolds will yield new contributions to $J$. This,
however, will not be the case between $Q$ and $\overline Q$. Evidence for
additional structure introduced by $Q$ and $\overline Q$ on the discrete space
was demonstrated when it became obvious that many new auxiliary terms were
arising due to the existence of $\xi$ and $\overline \xi$, which would
otherwise
vanish if we imposed $\xi={\overline \xi}$. ) The model dependency of the
scalar
potential is thus further emphasised by these considerations.

\section{Conclusion}

By utilizing the underlying mathematical approach of Connes construction we
have compactly described the BRS/anti-BRS structure of complicated models
involving multiple gauge and Higgs fields. Furthermore, we have proposed a
natural origin to the Zinn-Justin auxiliary scalar consistent with the
algebraic structure of noncommutative geometry. That noncommutative geometry
is indeed a natural arena for this type of investigation is emphasised
when it is recalled
that fermionic fields, having a canonical dimension of $3/2$, will not
contribute to the ``quantum term" of the action. The fermionic sector can, as
in the classical case, be introduced in a trivial way.

We have not explicitly considered the possibility of assymetric gauge choices
corresponding to breaking hermiticity as implied by (46). Actually, we note
that the
potential (53) implies that this hermiticity condition is already broken. Terms
do
arise in the calculation of
$\pi (\delta_{Q}\delta_{\overline Q}\Theta )_{\rm zero \; ghost \; no.}$ which
appear to fill this role, however these are not ${\overline S}_{X}$ invariant.
This
is actually a good result for two reasons:\newline
(i) such terms are associated with other terms for which zero ghost number
fails,
\newline
(ii) to explore the full range of possible gauges a new gauge parameter is
required,
implying the introduction of an additional ``quantum term".\newline
One possible such term which suggests itself is $\pi
(\delta_{Q}\tilde{\omega}\Theta)_{\rm zero\; ghost\; no.}$, allowing us to
maintain
the action as a functional of the curvature. This does not obviously introduce
BRS/anti-BRS invariant terms so that the naturalness of such an expression
needs to
be tested.

An interesting point which remains is the interpretation of $\beta$.
One natural possibility is that $\xi$
encodes information on the relative strengths of the group manifolds. That is,
we have no fixed notion of the relative sizes of these internal spaces. In this
sense $\xi$ contains information on the relative coupling strengths. A
topologically
more appealing possibility is that $\xi$ connects different choices of gauge on
the
different group manifolds.
The true
nature of this perhaps waits for a full quantum treatment. This is, of course,
speculation but signals the possibility of interesting new contributions for a
non-commutative quantum field theory.

\section{Acknowledgements}

BEH would like to thank R. Coquereaux for stimulating discussions and M.
Matsuda for the warm hospitality given
during his stay at the Aichi University of
Education where part of this work was completed. BEH would also like to thank
V. Sivananthan for many helpful discussions and valuable comments.

\end{document}